\documentclass[aps,prl,twocolumn,groupedaddress,showpacs,nofootinbib]{revtex4}
\usepackage{amsmath}
\usepackage{pstricks}
\begin{document}

\title{Almeida-Thouless transition below six dimensions}

\author{T.~Temesv\'ari}
\email{temtam@helios.elte.hu}
\affiliation{
Research Group for Theoretical Physics of the Hungarian Academy of Sciences,
%HAS Research Group for Theoretical Physics,
E\"otv\"os University, P\'azm\'any P\'eter s\'et\'any 1/A,
H-1117 Budapest, Hungary}

\date{\today}

\begin{abstract}
The existence of an Almeida-Thouless (AT) instability surface below the upper
critical dimension 6 is demonstrated in the generic replica symmetric field
theory. Renormalization flows from around the zero-field fixed point are
investigated. By introducing the temperature and magnetic field dependence
of the bare parameters, the fate of the AT line can be followed from mean
field ($d=\infty$) down to $d=6-\epsilon$.
\end{abstract}


\maketitle

Notwithstanding the relative simplicity of the relevant model postulated
by Edwards and Anderson \cite{EA}, the Ising spin glass problem has
resisted a thorough understanding for decades. Severe frustration makes
numerical simulations extremely hard and computer-time consuming, whereas
analytical methods must handle the inhomogenities caused by the quenched
disorder. The model was later extended and studied on the fully connected
lattice by Sherrington and Kirkpatrick (SK) \cite{SK}, the characterization
of the spin glass phase by the solution of Parisi (see Ref.\ \cite{MePaVi}
for a list of references) is now unanimously accepted as the true mean field
theory. This mean field spin glass proved to be very complicated, its
eqilibrium state breaking up to ultrametrically organized ergodic components,
commonly called pure states.
This complex phase space structure
survives in an external magnetic field up to a phase boundary called the
Almeida-Thouless line. Approaching this line from the paramagnetic
side an instability develops: using a replicated picture \cite{EA}, the
diverging spin glass susceptibility signals the break down of the replica
symmetric phase \cite{AT}, and replica symmetry breaking develops.

An alternative theory --- the so called droplet picture --- emerged, however,
and continues questioning the relevance of mean field ideas in finite
dimensional systems \cite{FiHu86}. In this theory the glassy phase
is much simpler, and is limited to zero field:
a convincing conclusion about the existence or lack of
an AT line may resolve a decades long debate about the structure of the
spin glass phase in the physical dimensions. Recent numerical simulations
\cite{YoungKatzgraber04,Jorgetal08} in three dimensions essentially
excluded the possibility of a transition in a field, whereas the four
dimensional case remains somewhat ambiguous (see \cite{MaNaZu98} for
references to earlier works). On the analytical side, we must mention
the scaling considerations in \cite{FiSo85} and renormalization group
(RG) calculations \cite{BrRo,PRLcikk}, whereas a leading order field
theoretical computation \cite{GrMoBr83} provided an AT line above 6 dimensions.
This letter tries to dissolve the misbelief that the AT line disappears
below the upper critical dimension, by explicitly calculating it close to,
but below $d=6$.

Ising spin glass transition in an external magnetic field can be studied
in the generic replica symmetric field theoretical model \cite{rscikk}
defined by the Lagrangean
$ \mathcal{L}=\mathcal{L}^{(2)}+\mathcal{L}^{\text{I}}$,
where
\begin{widetext}
%\[
\begin{equation}\label{L2}
\mathcal{L}^{(2)}=\frac{1}{2}\sum_{\mathbf p}\bigg[
\Big(\frac{1}{2} p^2+m_1\Big)\sum_{\alpha\beta}
\phi^{\alpha\beta}_{\mathbf p}\phi^{\alpha\beta}_{-\mathbf p}\\
+m_2\sum_{\alpha\beta\gamma}
\phi^{\alpha\gamma}_{\mathbf p}\phi^{\beta\gamma}_{-\mathbf p}+
m_3\sum_{\alpha\beta\gamma\delta}
\phi^{\alpha\beta}
_{\mathbf p}\phi^{\gamma\delta}_{-\mathbf p}
\bigg],
%\]
\end{equation}
and
\begin{multline}\label{LI}
\mathcal{L}^{\text{I}}=
-N^{\frac{1}{2}}\,h\,\sum_{\alpha\beta}
\phi^{\alpha\beta}_{{\mathbf p}=0}
-\frac{1}{6\sqrt{N}}
\sideset{}{'}\sum_{\mathbf {p_1p_2p_3}}
\bigg[
w_1\sum_{\alpha\beta\gamma}\phi^{\alpha\beta}
_{\mathbf p_1}\phi^{\beta\gamma}_{\mathbf p_2}
\phi^{\gamma\alpha}_{\mathbf p_3}+w_2\sum_{\alpha\beta}\phi^{\alpha\beta}
_{\mathbf p_1}\phi^{\alpha\beta}
_{\mathbf p_2}\phi^{\alpha\beta}_{\mathbf p_3}%\\[5pt]
+w_3\sum_{\alpha\beta\gamma}\phi^{\alpha\beta}
_{\mathbf p_1}\phi^{\alpha\beta}
_{\mathbf p_2}\phi^{\alpha\gamma}_{\mathbf p_3}\\
+w_4\sum_{\alpha\beta\gamma\delta}\phi^{\alpha\beta}
_{\mathbf p_1}\phi^{\alpha\beta}
_{\mathbf p_2}\phi^{\gamma\delta}_{\mathbf p_3}
+w_5\sum_{\alpha\beta\gamma\delta}\phi^{\alpha\beta}
_{\mathbf p_1}\phi^{\alpha\gamma}
_{\mathbf p_2}\phi^{\beta\delta}_{\mathbf p_3}%\\[2pt]
+w_6\sum_{\alpha\beta\gamma\delta}\phi^{\alpha\beta}
_{\mathbf p_1}\phi^{\alpha\gamma}
_{\mathbf p_2}\phi^{\alpha\delta}_{\mathbf p_3}
+w_7\sum_{\alpha\beta\gamma\delta\mu}\phi^{\alpha\gamma}
_{\mathbf p_1}\phi^{\beta\gamma}
_{\mathbf p_2}\phi^{\delta\mu}_{\mathbf p_3}
+w_8\sum_{\alpha\beta\gamma\delta\mu\nu}\phi^{\alpha\beta}
_{\mathbf p_1}\phi^{\gamma\delta}
_{\mathbf p_2}\phi^{\mu\nu}_{\mathbf p_3}
\bigg].
\end{multline}
\end{widetext}
The $n(n-1)/2$ component fields are symmetric in the replica indices and
$\phi^{\alpha\alpha}\equiv 0$; the spin glass limit requires $n\to 0$.
The total number of spins $N$ is included to ensure the correct thermodynamic
limit, and momentum conservation is understood in the primed summation.
The zero-field paramagnetic phase corresponds to higher symmetry
\cite{droplet}, with all the bare parameters but $m_1$ and $w_1$ zero,
it has a unique mass $\Gamma$, and below the upper critical dimension
$d_u=6$ the spin glass transition is governed by the fixed point 
\cite{HaLuCh76} $w_1^{*2}\equiv w^{*2}=\epsilon/(2-n)$
and $m_1^*=-\epsilon/2$,
$\epsilon=6-d$.
%The AT transition is a kind of quadratic symmetry breaking,
%its study becomes clearer by the introduction of the nonlinear scaling
%fields \cite{Wegner} satisfying the {\em \/exact} renormalization flows
%$\dot{g_i}=\lambda_i\,g_i$.
The mass is split by an external magnetic field into the three different
components $\Gamma_{\text{R}}$, $\Gamma_{\text{A}}$ and $\Gamma_{\text{L}}$,
thus generating --- while replica symmetry is still preserved --- a kind
of quadratic symmetry breaking. The crossover region is best studied
by the introduction of the nonlinear scaling
fields \cite{Wegner} satisfying the {\em \/exact} renormalization flows
$\dot{g_i}=\lambda_i\,g_i$.
[The $\lambda_i$'s of the mass sector ($i=1,2,3$) were computed from the
renormalization flow equations, in leading order,
in Ref.\ \cite{Iveta}.] %, the field-like one can also be easily
%calculated providing $\lambda_0=4-\epsilon/3$.
The RG equations provide a
way to express the bare parameters of the Lagrangean in terms of the scaling
fields, hence the masses can be computed as functions of the $g_i$'s:
\begin{align}
\Gamma_{\text{R}}&=g_1+2g_2+g_3+O(\epsilon),\notag\\
\Gamma_{\text{A}}&=g_1-(n-4)g_2-(n-3)g_3+O(\epsilon),\label{masses}\\
\Gamma_{\text{L}}&=g_1-2(n-2)g_2+\frac{(n-2)(n-3)}{2}g_3+O(\epsilon).\notag
\end{align}
The $O(\epsilon)$ terms neglected above have two contributions: the one-loop
self-energy (which is computable at that order) and corrections to the bare
masses expressed in terms of the $g_i$'s --- this, however, is not available
in a leading order RG calculation. Coupling-like scaling fields $g_i$'s
with $i>3$ enter also at this  $O(\epsilon)$ level.
Eq.\ (\ref{masses}) can be derived by
fixing the bare parameters such that $\langle\phi^{\alpha\beta}_{\mathbf p}
\rangle\equiv 0$; %this can be achieved
%by tuning $u_0$.
this condition determines $g_0$ unambigously in terms of the other $g_i$'s.
Two critical surfaces can be found from Eqs.\ (\ref{masses}) in the low
temperature ($g_1<0$) regime:
\begin{itemize}
\item $\Gamma_{\text{R}}=0$, $\Gamma_{\text{A}}$ and $\Gamma_{\text{L}}$
both positive --- i.e.\ an Almeida-Thouless instability --- for $g_2<-g_1$;
\item $\Gamma_{\text{A}}=0$, $\Gamma_{\text{L}}$ and $\Gamma_{\text{R}}$
positive ($n\gtrsim 0$) for $g_2>-g_1$.
\end{itemize}
The common boundary of these two manifolds (which are two-dimensional now,
but allowing for coupling-like scaling fields $g_i$, $i>3$, they will have
a complicated higher dimensional structure) for $g_2=-g_1$ is massive only
in the longitudinal sector \cite{PRLcikk}.

We are now interested in the RG flows along the AT instability surface
when starting in the crossover region. The first order RG equations were
all presented in Ref.\ \cite{Iveta}, their structure is best displayed by
the following (temporary) redefinition of the couplings:
$w_i/\sqrt{\epsilon}\,\to\,
w_i$, which are now, like the masses, order unity. With the scaling factor
$e^{dl}$, and $t\equiv \epsilon l$:
\begin{align}
\frac{dm_i}{dl}&=2m_i-\epsilon\mathcal M_i(m_1,m_2,m_3;w_1,\dots,w_8),
 \,\, i=1,2,3; \label{mass_flow}\\
\frac{dw_i}{dt}&=\frac{1}{2}w_i+\mathcal W_i(m_1,m_2,m_3;w_1,\dots,w_8),
\,\,\,\, i=1,\dots,8. \label{coupling_flow}
\end{align}
The $\mathcal M_i$ and $\mathcal W_i$ functions are quadratic and cubic,
respectively, in the couplings.
%The initial conditions required for the crossover process from around
%the zero-field fixed point follow from the transformation formulae between
%bare parameters and nonlinear scaling fields shown in Table \ref{table}.
%\begin{align*}
%m_1&=   \\
%m_2&=   \\
%m_3&=   \\
%\hline \\
%w_1 &=   \\w_2 &=   \\
%w_3 &=   \\w_4 &=   \\
%w_5 &=   \\w_6 &=   \\
%w_7 &=   \\w_8 &=   
%\end{align*}
The most important feature of the RG equations above is that the flow
parameter $l$ in the mass sector, Eq.\ (\ref{mass_flow}), is much larger
for $\epsilon\ll 1$ than $t$ of the couplings, Eq.\ (\ref{coupling_flow}).
Thus the masses renormalize in the background of the adiabatically
slow couplings: the anomalous (A) and longitudinal (L) components,
as they are $O(1)$ on the AT surface, blow up, whereas the replicon (R) one,
$2m_1=O(\epsilon)$, evolves into its adiabatic fixed point determined by the initial values of the couplings $w_1^{0+}$ and $w_2^{0+}$. While $w_1^{0+}
=w_1(t=0)=w_1^*$, we must carefully follow the development of $w_2$ in the
transient regime from $w_2(t=0)=0$
\footnote{More precisely, $w_2(t=0)\ll \epsilon$ in the physically relevant part of
the AT surface.}
to $w_2^{0+}\equiv w_2(t)$ with
$\epsilon\ll t\ll 1$ for the following reason:
For $l\gg 1$, i.e.\ $t\gg \epsilon$, $w_1$ and $w_2$ decouple from the other
bare parameters, and their
flow can be put into the pair of equations:
\begin{align}
\frac{dw_1}{dt}&=\frac{1}{2}w_1+g_n(r)\,w_1^3\notag \\
\frac{dr}{dt}&=-h_n(r)\,w_1^2, \notag
\end{align}
where $g_n(r)$ and $h_n(r)$ are cubic and quartic polynomials of $r$ with
coefficients which are simple polynomials of $n$, and $r\equiv w_2/w_1$.
For the case $n=0$, these equations were derived and discussed in Ref.\ \cite{BrRo}.
We are now interested in the more generic case $0\le n\lesssim \epsilon$, and observe that the 
qualitative behaviour of the renormalization flow changes drastically when the
initial value of $r(0)=w_2^{0+}/w_1^{0+}$ passes through $r_1^*=\frac{3}{10}n+O(n^2)$,
the unstable fixed point $r_1^*$ being the solution of the equation $h_n(r)=0$.
For $r(0)>r_1^*$, we have runaway trajectories already noticed in \cite{BrRo} with
$w_1\to \infty$ and $r\to r_2^*\cong 14.4+O(n)$; whereas for $r(0)<r_1^*$, $w_2$
immediately becomes negative, which
%--- as we will argue later ---
is physically
nonsense.

To get $r(0)$, we must integrate Eq.\ (\ref{coupling_flow}) for $i=2$ in
the transient regime from $t=0$ to $\epsilon\ll t \ll 1$, thereby eliminating
nonreplicon modes in the process of hardening anomalous and longitudinal
masses. This is feasible using Eq.\ (65) of Ref.\ \cite{Iveta} together with the
table between the different sets of couplings in Eq.\ (49) of
Ref.\ \cite{rscikk}, resulting in $r(0)<r_1^*$ for $0<n\ll 1$ and starting
close enough to the zero-field fixed point, whereas the spin glass case
is exceptional with the condition $r(0)>r_1^*$ {\em always\/} fulfilled.
%\[
%r(0)=3\,\epsilon,\quad 0\le n\lesssim \epsilon \ll 1.
%\]
%In the spin glass domain $n\to 0$, we always have $r_1^*<r(0)$, and runaway
%trajectories with $w_1$ and $w_2$ blowing up --- with fixed ratio ---
%begin to develop.

Runaway flows along critical surfaces have been associated with first
order transitions in some common situations with crossover phenomena
\cite{Amit}, and although this scenario cannot be ruled out completely
for the spin glass either, we will argue that renormalization of the bare
couplings on the AT surface towards their {\em low-temperature} limit may cause
the runaway trajectories in this renormalization scheme. To see this, we recall
the derivation of the microscopic Lagrangean in Ref.\ \cite{rscikk}, and the
necessity to redefine the fields as $c\,\phi^{\alpha\beta}_{\mathbf p}
\rightarrow \phi^{\alpha\beta}_{\mathbf p}$, with $c\sim T$, to ensure the
proper normalization of the kinetic term in $\mathcal{L}^{(2)}$. This will
cause the couplings diverge even if they disappeared for $T\to 0$ otherwise.
That kind of normalization was essential in the derivation of Eqs.\
(\ref{mass_flow}) and (\ref{coupling_flow}), manifested in the flowing
$\eta$ exponents of the three different mass modes. As our approximate RG
equations are valid only for $w_i=O(1)$, 
%for detecting the
%proper zero-temperature
%behaviour in this small $\epsilon$ regime on the AT surface, one
%probably needs to
%modify the RG scheme in a way similar to that in Ref.\ \cite{Lawrie} for
%the coexistence-line in the isotropic $n$-vector model.
one probably needs to modify the RG scheme for detecting the proper
zero-temperature behavior
on the AT surface in this small $\epsilon$ regime.
This is, however,
out of the scope of the present work.

In the remaining part of this letter we want to locate the AT-line of
the original Edwards-Anderson spin glass model on the AT-surface of
the field theory above. For this reason, we must find out the dependence
of the bare parameters in Eqs.\ (\ref{L2}) and (\ref{LI}) on temperature
($T$) and magnetic field ($H$). The criterium which is adopted here is
that the tree approximation of the field theory (i.e.\ neglecting loops)
be equivalent with the accepted mean field theory of the Ising spin glass,
the SK model, whose replicated partition function has the form \cite{SK}:
\begin{equation}\label{SK1}
\overline{Z^n}\sim% \bigg[\prod_{  \alpha<\beta
                   %                }\int
\int\mathcal{D}q		   %\sqrt{\frac{N}{2\pi}}\,(kT)\,	 dq_{\alpha\beta}\bigg]
\,\exp\bigg\{-N\Big[\frac{(kT)^2}{2J^2}\sum_{\alpha<\beta}q_{\alpha\beta}^2
-\ln \zeta\Big]\bigg\},			
\end{equation}
where
\begin{equation}\label{SK2}
\zeta =\underset{\{S^{\alpha}\}}{\mathrm {Tr}}
\exp\Big(\sum_{\alpha<\beta}q_{\alpha\beta}\,S^{\alpha}S^{\beta}+\frac{H}{kT}
\sum_{\alpha}\,S^{\alpha}\Big),
\end{equation}
$\int\mathcal{D}q\equiv %\big[
\prod\limits_{  \alpha<\beta
                                   }\big(\int
		   N^{\frac{1}{2}}\frac{kT}{\sqrt{2\pi J^2}}\,        dq_{\alpha\beta}\big)
$ and $J^2$, the variation of the Gaussian distribution of the random Ising
interactions, sets the energy scale. 
In the tree approximation fluctuations are omitted, which can be achieved by setting
$\phi^{\alpha\beta}_{\mathbf p=0}=\sqrt{N}\,q_{\alpha\beta}$ and zero for 
$\phi^{\alpha\beta}_{\mathbf p\not=0}$ in (\ref{L2}) and (\ref{LI}), and comparing it
with (\ref{SK1}) and (\ref{SK2}). Not forgetting that the bare parameters of the
field theory are finally tuned by the transformation $\phi^{\alpha\beta}_{\mathbf p}
-\sqrt{N}\,q\,\delta^{\text{Kr}}_{\mathbf p=0} \rightarrow 
\phi^{\alpha\beta}_{\mathbf p}$ --- rendering the one-point function to zero ---,
where $q$ is the {\em exact\/} replica symmetric order parameter, they are expressed
by $T$ and $H$ in the vicinity of the mean field critical point
$kT_c^{\text{mf}}=J$ as follows:
\begin{align}
%wh&=\frac{1}{2}\,\Big(\frac{H}{kT}\Big)^2
wh&=\frac{1}{2}(H/kT)^2
-(m_{1c}-\tau)(wq)+\frac{1}{2}(n-2)
(wq)^2+\dots \notag\\
m_1&=(m_{1c}-\tau)+(wq)+%\dots \notag\\
(H/kT)^2%\Big(\frac{H}{kT}\Big)^2
-\frac{1}{2}\,q^2\,\Big[u_{01}+u_{02}+\notag\\
&\mathrel{\phantom{=}}\frac{1}{3}(n-1)
\,u_{03}+\frac{1}{3}n(n-1)\,u_{04}\Big]\dots \label{bare_parameters}\\
m_2&=-(wq)%+\dots \notag\\
-(H/kT)^2%-\Big(\frac{H}{kT}\Big)^2
-\frac{1}{3}\,q^2\,\big[(n-3)\,u_{01}+u_{03}\big]
+\dots \notag\\
m_3&=%O(q^2,qH^2)\notag\\
-\frac{1}{6}\,q^2\,\big[u_{01}+2u_{04}\big]+\dots \notag\\
w_1&=w+O(q,H^2),\quad\text{and}\quad w_i=O(q,H^2)\quad i=2\dots 8.\notag
\end{align}
Neglected terms above are higher orders in $q$ and $H^2$. The quartic couplings are
not included here, although their calculation is similarly straightforward, and
they may be inportant above 8 dimensions. $\tau>0$ measures the distance from
the critical temperature of the field theory ($T_c$), whereas $m_{1c}=
-\frac{k^2}{2J^2}\,(T_c^{\text{mf}\,2}-T_c^2)$ gives the shift in the critical
temperature, and is therefore one-loop order.
The field theory is defined by $\tau$, $(H/kT)^2$ and by the bare parameters
of the symmetrical theory (zero magnetic field paramagnet): $w$
(cubic coupling), $u_{01}$, $u_{02}$, $u_{03}$, $u_{04}$ (quartic couplings),
\dots etc.\ (see \cite{nucl} for the classification of the quartic couplings).
To reproduce the SK model results in the tree approximation, we must put
$w=1$, $u_{01}=3$, $u_{02}=2$, $u_{03}=-6$ and $u_{04}=0$.

The condition $\langle\phi^{\alpha\beta}_{\mathbf p}
\rangle= 0$ provides us the equation of state, i.e.\ the order parameter
$q$ around $T_c$; it is used here to eliminate $\tau$ from our results,
replacing it by $q$.
%The structure of this equation is shown schematically
%in Fig.\ \ref{equation_of_state}.
The calculation of the one-loop
contribution to the equation of state, and to the replicon mass
%in
%Fig.\ \ref{replicon_mass},
is somewhat lengthy due to the complicated replica
structure even in the case of replica symmetry. Nevertheless, it is still
feasible by the methods of Ref.\ \cite{rscikk}.
%\begin{figure}[h]\caption{Equation of state for expressing $\tau$ in terms
%of $q$.}\label{equation_of_state}
%\begin{pspicture}(8,4)%\psgrid
%\rput[Br](3,2){\large $0=2\,h\quad+$} \rput[Bl](5,2){\large $+\dots$}
%\pscircle(4,2.6){0.6}
%\psline[linestyle=dotted](4,1.4)(4,2)
%\end{pspicture}
%\end{figure}
%
%The one-loop calculation of the replicon mass $\Gamma_{\text{R}}$ is
%schematically displayed in Fig.\ \ref{replicon_mass}.
%\begin{figure}[h]\caption{The replicon mass $\Gamma_{\text{R}}$}
%\label{replicon_mass}
%\begin{pspicture}(8.5,4)%\psgrid
%\rput[Br](2.5,2){\large $\Gamma_{\text{R}}=2\,m_1+$}
%\psellipse(4,2.08)(1,0.6)\psline[linestyle=dotted](2.7,2.08)(3,2.08)
%\psline[linestyle=dotted](5,2.08)(5.3,2.08)
%\rput[Bl](5.5,2){\large $+$}%\pscurve(6.9,2.08)(6.5,2.8)(7.3,2.8)(6.9,2.08)
%\pscircle(6.7,2.68){0.6}
%\psline[linestyle=dotted](6.1,1.4)(6.7,2.08)
%\psline[linestyle=dotted](7.3,1.4)(6.7,2.08)
%\rput[Bl](7.5,2){\large $+\dots$}
%\end{pspicture}
%\end{figure}
The result valid for
$d>8$, including the SK model by simply taking $d=\infty$, can be put into
the scaling form
\begin{equation}\label{scaling1}
\Gamma_{\text{R}}=(wq)^2\,\tilde\Gamma_{\text{R}}(x,y),\quad x\equiv
%(\frac{H}{kT}\Big)^2/(wq)^3 \quad\text{and}\quad y\equiv n/(wq).
\frac{(H/kT)^2}{(wq)^3} \quad\text{and}\quad y\equiv \frac{n}{(wq)}.
\end{equation}
The scaling function $\tilde\Gamma_{\text{R}}$ has the simple linear form
\begin{equation}\label{scaling_function}
\tilde\Gamma_{\text{R}}(x,y)=ax+by+c,\qquad d>6,
\end{equation}
with $a$ and $b$ analytical down to 4 and 6 dimensions,
respectively, %whereas $c=-16\,w^2\,\frac{1}{N}\sum\frac{1}{p^8}+\dots$. 
and having their loop-expansions in terms of $I_k\equiv
\frac{1}{N}\sum^{\Lambda}\frac{1}{p^k}$:
\begin{align}\notag%\label{a_and_b}
a&=1-2w^2 I_4\quad \text{and}\quad\\\label{a_and_b}\\
b&=1-2w^2 I_6+(-u_{10}+u_{30}+4u_{40})I_4;
\quad d>6.\notag
\end{align}
$c$ however blows up at 8 dimensions, due to the infrared divergence
developing in the first order contribution behind the mean field term:
\begin{equation}\label{c}
c=-\frac{2}{3}u_{20}w^{-2}-16w^2I_8
+\text{terms with $I_6$ and $I_4$}, \quad d>8.
\end{equation}
As a result, scaling of the replicon mass turns to the following form
when $6<d<8$:
\begin{gather}%\label{scaling2}
\Gamma_{\text{R}}=(wq)^{d/2-2}\,\tilde\Gamma_{\text{R}}(x,y),\notag\\%[3pt]
\label{scaling2}\\\notag
\text{with}\quad x\equiv
\frac{(H/kT)^2}{(wq)^{d/2-1}} \quad\text{and}\quad y\equiv
\frac{n}{(wq)^{d/2-3}}.
\end{gather}
The scaling function preserves the form (\ref{scaling_function}) with
$a$ and $b$ in (\ref{a_and_b}), the constant $c$ however becomes,
instead of (\ref{c}):
\begin{equation}\label{cvar}
c=-16w^2\int^{\infty} \frac{d^dp}{(2\pi)^d}\,\frac{1}{p^4(p^2+2)^2},
\quad 6<d<8.
\end{equation}

The zeros of the scaling function provide the AT transitions, and two
important cases can be studied for $d>6$:
\begin{itemize}
\item $H=0$, i.e.\ $x=0$ and $y=-c/b$. This case has been discussed in
\cite{nucl}.
\item The spin glass limit $n=0$, i.e.\ $y=0$ and $x=x_0=-c/a$. The AT
line close to $T_c$ in the two regimes is:
\begin{equation}\label{d>6AT}
\begin{cases}
(H/kT)^2=x_0\, (wq)^3 & 8<d, \\
(H/kT)^2=x_0\, (wq)^{d/2-1} & 6<d<8.
\end{cases}
\end{equation}
\end{itemize}
From (\ref{a_and_b}) and (\ref{c}), with $u_{20}=2$, the SK value $x_0=4/3$
is reproduced, whereas $x_0$ becomes one-loop order for $6<d<8$, see
(\ref{cvar}).

Below 6 dimensions, the leading scaling behaviour can be obtained by
using fixed point values in (\ref{bare_parameters}), and also neglecting
correction terms providing:
\begin{align}
w^*h&=\frac{1}{2}(H/kT)^2
%+\tau\,(w^*q)+\frac{1}{2}(n-2)\,
-(m_{1}^*-\tau)\,(w^*q)+\frac{1}{2}(n-2)\,
(w^*q)^2, \notag\\
m_1&=m_{1}^*-\tau+(w^*q),\quad m_2=-(w^*q), \quad m_3=0, \notag\\[4pt]
w_1&=w^*, \qquad\qquad\quad w_i=0,\qquad\quad i=2\dots 8.\notag
\end{align}
After eliminating $\tau$ by the equation of state, we are left with a
two-parameter theory, with the simple RG flows close to the fixed point:
\begin{equation}\label{q_H_scaling}
\dot q \simeq (2-\epsilon/2+\eta^*/2)\,q\quad \text{and}\quad
\dot{(H/kT)^2} \simeq \lambda_0\, (H/kT)^2,
%\quad \text{with} \quad \lambda_0=4-\epsilon/2+\eta^*/2.
\end{equation}
with $\lambda_0=4-\epsilon/2-\eta^*/2$ and $\eta^*=-\epsilon/3$.
The scaling fields can now be expressed as
\begin{gather}
g_i\simeq(w^*q)^{z_i}\,\tilde g_i(x),\notag\\[3pt]
x= \frac{(H/kT)^2}{(w^*q)^\delta}\quad\text{and}\quad
\delta=\frac{4-\epsilon/2-\eta^*/2}{2-\epsilon/2+\eta^*/2}.\notag
\end{gather}
From (\ref{q_H_scaling}) follows that $x$ is invariant under renormalization,
and $z_i$ must be $(2-\epsilon/2+\eta^*/2)^{-1}\,\lambda_i$ for $\dot{g_i}
=\lambda_i\,g_i$ be satisfied. Any observable $\mathcal O$ satisfying the
approximate RG flows $\dot{\mathcal O}\simeq k_{\mathcal O}\, {\mathcal O}$
around the fixed point can now be written as ${\mathcal O}\sim
(w^*q)^{k_{\mathcal O}/(2-\epsilon/2+\eta^*/2)}$ times a function of $x$.
For a mass $k=2-\eta^*=(\delta-1)\,(2-\epsilon/2+\eta^*/2)$, and therefore
the replicon mass takes the scaling form
\begin{equation}\label{scaling3}
\Gamma_{\text{R}}\simeq (w^*q)^{\delta-1}\,\tilde \Gamma_{\text{R}}(x),
\quad x= \frac{(H/kT)^2}{(w^*q)^\delta}.
\end{equation}
The most important new feature of (\ref{scaling3}) when compared with the
$d>6$ cases, Eqs.\ (\ref{scaling1}) and (\ref{scaling2}), is the lack of
the second scaling variable, which is proportional to $n$. The AT line
ends now in the zero field critical point
\footnote{\label{fnote}When $6<d<8$ the AT line takes the form $(H/kT)^2\sim n^{1-2/\epsilon}\,
(q-q_c)$ for $n\gtrsim 0$, where $q_c\sim n^{-2/\epsilon}$.
For $d\geq 8$ the mean field phase diagram restores
\cite{FiSo85,rscikk,nucl}.}
even for $n$ small but nonzero:
\begin{equation}\label{d<6AT}
(H/kT)^2=x_0\,(w^*q)^\delta, \quad x_0=-n+[2+O(n)]\epsilon+O(\epsilon^2),
\end{equation}
and it disappears completely for $n>2\epsilon$.
\begin{figure}\caption{AT lines on the two sides of the upper
critical dimension 6. See (\ref{d>6AT}), (\ref{d<6AT}) and footnote
\ref{fnote}.}\label{AT_lines}
\vspace{10pt}
\begin{pspicture}(0,-1)(8,4)%\psgrid
\psline[linewidth=0.4pt]{->}(5,0)(8,0)\uput[r](8,0){$q$}
\psline[linewidth=0.4pt]{->}(5,0)(5,4)\uput[r](5,3.9){$H^2$}
\pscurve(5,0)(6,0.2)(7,1)(8,3.7)%\parabola(7.5,4)(4.5,0)
\psline[linewidth=0.4pt]{->}(0.5,0)(4,0)\uput[r](4,0){$q$}
\psline[linewidth=0.4pt]{->}(0.5,0)(0.5,4)\uput[r](0.5,3.9){$H^2$}
\pscurve(0.5,0)(1.5,0.4)(2.5,2.6)(2.75,3.7)
\pscurve(2,0)(3,0.2)(4,1)\uput[d](2,0){$q_{c}$}
\psline[linewidth=0.4pt](2,-0.1)(2,0.1)
\uput[r](1,3){\psframebox[linewidth=0.2pt,framesep=2pt]{$n=0$}}
\uput[r](2.4,1){\psframebox[linewidth=0.2pt,framesep=2pt]{$n\gtrsim 0$}}
\uput[r](5.5,3){\psframebox[linewidth=0.2pt,framesep=2pt]{$0\le n<2\epsilon$}}
\rput[l](1.3,-0.9){(a) $d\gtrsim 6$}\rput[l](5.8,-0.9){(b) $d\lesssim 6$}
\rput(3.28,2.5){$\sim q^{d/2-1}$}
\rput(6.5,1){$\sim q^{\delta}$}
\end{pspicture}
\end{figure}
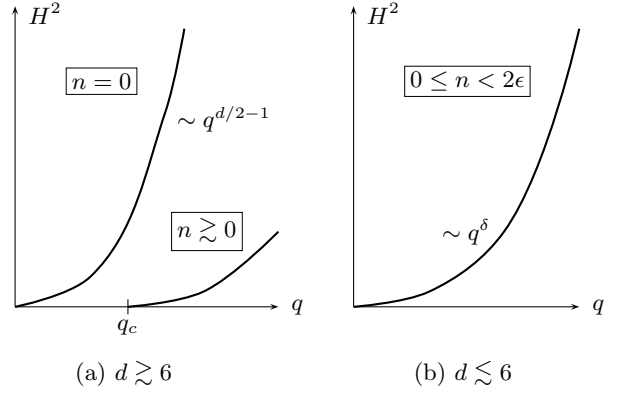

To conclude, we followed the fate of the AT line from mean field down to
$d=6-\epsilon$.
%Runaway flows towards zero temperature behavior start
%from the whole transition surface exceptionally in the spin glass case
%($n=0$).
An exceptional feature of the spin glass case ($n=0$) is that
the runaway flows towards zero-temperature behavior --- found
below $d=6$ --- originate in the close vicinity of the zero-field fixed point.
Our results do not exclude a possible lack of the AT surface
in $d=3$ --- as suggested by recent numerical works
\cite{YoungKatzgraber04,Jorgetal08} ---, a scenario, with some lower
critical dimension to explain this, has been suggested in \cite{nucl}.

\bibliography{spinglass}

\end{document}